# Modeling Two First Hydrostatic Core Candidates Barnard 1b-N and 1b-S

Hao-Yuan Duan (段皓元)[1,2] , Shih-Ping Lai (賴詩萍)[1,2,3] , Naomi Hirano (平野尚美)[4] , and Travis J. Thieme (陸哲軒)[1,2]
[1] Institute of Astronomy, National Tsing Hua University, No. 101, Section 2, Kuang-Fu Road, Hsinchu 30013, Taiwan; slai@phys.nthu.edu.tw
[2] Center for Informatics and Computation in Astronomy, National Tsing Hua University, No. 101, Section 2, Kuang-Fu Road, Hsinchu 30013, Taiwan
[3] Department of Physics, National Tsing Hua University, No. 101, Section 2, Kuang-Fu Road, Hsinchu 30013, Taiwan
[4] Academia Sinica Institute of Astronomy and Astrophysics, 11F of Astronomy-Mathematics Building, AS/NTU, No.1, Sec. 4, Roosevelt Rd, Taipei 10617, Taiwan
Received 2021 August 2; revised 2023 January 18; accepted 2023 January 19; published 2023 April 19

## Abstract

A first hydrostatic core (FHC) is proposed to form after the initial collapse of a prestellar core, as a seed of a Class 0 protostar. FHCs are difficult to observe because they are small, compact, embedded, and short lived. In this work, we explored the physical properties of two well-known FHC candidates, B1-bN and B1-bS, by comparing interferometric data from Submillimeter Array (SMA) 1.1 and 1.3 mm and Atacama Large Millimeter/submillimeter Array (ALMA) 870 $\mu$m observations with simulated synthesis images of the two sources. The simulated images are based on a simple model containing a single, hot compact first-core-like component at the center surrounded by a large-scale, cold and dusty envelope described by a broken power-law density distribution with an index, $\alpha$. Our results show that the hot compact components of B1-bN and B1-bS can be described by temperatures of $\sim$500 K with a size of $\sim$4 au, which are in agreement with theoretical predictions of an FHC. If the $\alpha$ inside the broken radii is fixed to $-1.5$, we find $\alpha \sim -2.9$ and $\sim -3.3$ outside the broken radii for B1-bN and B1-bS, respectively, consistent with theoretical calculations of a collapsing, bounded envelope and previous observations. Comparing the density and temperature profiles of the two sources with radiation-hydrodynamic simulations of an FHC, we find both sources lie close to, but before, the second collapse stage. We suggest that B1-bS may have started the collapsing process earlier compared to B1-bN, since a larger discontinuity point is found in its density profile.

*Unified Astronomy Thesaurus concepts:* Star formation (1569)

## 1. Introduction

Probing the collapsing process from a prestellar core to a Class 0 protostar is key to understanding the earliest stage of low-mass star formation. Larson (1969) first proposed that this initial isothermal collapse (the first collapse) would produce a hydrostatic equilibrium object, known as a first hydrostatic core (FHC), at the center of the collapsing cloud. Masunaga et al. (1998) performed radiation-hydrodynamic simulations of FHCs, showing that they have a size of $\sim$5 au and a mass of $\sim$0.05 $M_\odot$ for an initial temperature of 10 K and a typical opacity of $\sim$0.01 cm$^2$ g$^{-1}$ at a wavelength of 10 $\mu$m, which are independent of the parent cloud's mass and initial density distribution. As an FHC grows, the central density increases from $10^{-16}$ to $10^{-10}$ g cm$^{-3}$, and the collapse becomes adiabatic due to high optical depth (Inutsuka 2012). The core can then be heated to over 100 K in a very short period ($\sim$5000 yr) due to accretion shocks (Masunaga & Inutsuka 2000). An FHC has a short lifetime of $10^3$–$10^4$ yr (Boss & Yorke 1995; Masunaga & Inutsuka 2000) before the temperature becomes high enough ($\sim$2000 K) to dissociate hydrogen molecules and trigger the second collapse to form a protostar, which is observed as a Class 0 object (Andre et al. 1993). MHD simulations suggest that outflows can be launched as early as the FHC phase, and these outflows have a wide opening angle and slow speed ($\sim$5 km s$^{-1}$) due to FHCs having a more shallow gravitational potential compared to a Class 0 protostar (Machida et al. 2008; Price et al. 2012).

FHCs are the best targets for studying what happens at the very beginning of low-mass star formation. However, not surprisingly, FHCs are extremely difficult to detect directly because they are compact, embedded, and short lived. Thus, only candidate FHCs have been proposed until now, but none of them have been confirmed. Observationally, FHC candidates are detected as a faint point source in the far-infrared ($>$70 $\mu$m) to millimeter continuum emission, with no observable emission below 30 $\mu$m: B1-bN and B1-bS (Pezzuto et al. 2012; Hirano & Liu 2014), Chamaeleon-MMS1 (Belloche et al. 2006; Väisälä et al. 2014), L1448 IRS2E (Chen et al. 2010), Per-bolo 58 (Enoch et al. 2010; Dunham et al. 2011), L1451 mm (Pineda et al. 2011; Maureira et al. 2017), N6 and SM1N in Ophiuchus (Friesen et al. 2018), and MC 35 mm (Fujishiro et al. 2020). These candidates are characterized by very cold spectral energy distributions (SEDs), with a range of $\sim$10–30 K, very low luminosities ($<$0.1–0.25 $L_\odot$), and most of them are found to have slow, poorly collimated outflows.

Among the candidates mentioned above, B1-bN and B1-bS are two well-known extremely young low-mass protostellar objects separated by $\sim$20″. They were discovered by Hirano et al. (1999) in the dense molecular cloud core Barnard 1-b (B1-b) within the Perseus molecular cloud complex. Pezzuto et al. (2012) have detected far-infrared emission from B1-bN and B1-bS using the Herschel Space Observatory and presented SED fitting results of these two sources. They showed that the emission from each source roughly corresponds to a compact central object with a surrounding dusty envelope, and proposed that these sources are candidates for FHCs. The distance to the B1-b core was recently measured to be $301 \pm 4$ pc using a combination of stellar photometry, astrometric data, and $^{12}$CO spectral-line maps (Zucker et al. 2018).

**Table 1**
The SMA Observations—Continuum

| $\lambda$ ($\mu$m) | Configuration | Beam Size | Flux Density (bN) (Jy) | Flux Density (bS) (Jy) | Error[a] (Jy) | rms (mJy beam$^{-1}$) |
|---|---|---|---|---|---|---|
| 1300 | Compact + subcompact | $3\rlap{.}''6 \times 2\rlap{.}''7$ | $1.85 \times 10^{-1}$ | $3.08 \times 10^{-1}$ | $6.52 \times 10^{-2}/8.34 \times 10^{-2}$ | 1.1/1.1 |
| 1100 | Subcompact | $4\rlap{.}''8 \times 2\rlap{.}''5$ | $2.88 \times 10^{-1}$ | $4.16 \times 10^{-1}$ | $6.61 \times 10^{-2}/7.90 \times 10^{-2}$ | 3.1/3.1 |
| 870 | Compact | $2\rlap{.}''0 \times 1\rlap{.}''7$ | $3.95 \times 10^{-1}$ | $8.87 \times 10^{-1}$ | $1.03 \times 10^{-1}/1.54 \times 10^{-1}$ | 2.6/2.6 |

**Note.**
[a] Error estimated from two-dimensional Gaussian fittings to the sources.

Recently, many interferometric molecular line observations and kinematic studies have been carried out toward the two sources. Huang & Hirano (2013) derived the ratio of $N_2D^+/N_2H^+$ from the Submillimeter Array (SMA; Ho et al. 2004) observations in both B1-bN and B1-bS to be ~0.2, which is comparable to those of prestellar cores in the late evolutionary stage or Class 0 protostars in the early evolutionary stage. They also suggest that B1-bN is in an earlier evolutionary stage compared to B1-bS because the nondetection of $H^{13}CO+$ toward B1-bN suggested the significant CO depletion in this source. Hirano & Liu (2014) found CO outflows associated with B1-bN and B1-bS using the SMA, while Gerin et al. (2015) detected $CH_3OH$ and $H_2CO$ outflows for both sources using the Plateau de Bure Interferometer (PdBI). These results showed that the maximum velocities of the B1-bN and B1-bS outflows are 3–5 and 7–8 km s$^{-1}$, respectively, suggesting B1-bS is slightly more evolved, due to the higher velocities. Evidence of first-core-like disks associated with B1-bN and B1-bS was found by modeling the high-resolution submillimeter continuum emissions from the Atacama Large Millimeter/submillimeter Array (ALMA; Gerin et al. 2017). Fuente et al. (2017) revealed the rotation of a pseudo-disk with $NH_2D$ observations from the NOrthern Extended Millimeter Array (NOEMA). Marcelino et al. (2018) detected rich complex organic molecules in B1-bS with ALMA, and showed that B1-bS is a hot corino, though no such hot corino signature is found in B1-bN. Recent studies have started to reveal hot corinos around low-mass Class 0 protostars with ALMA. More observations are needed to show whether hot corinos can also be associated with first cores.

The above recent studies could suggest that B1-bN and B1-bS may have passed the first core stage already, while still being in an earlier evolutionary stage than most of the known Class 0 protostars. In this paper, we aim to explore the physical properties of B1-bN and B1-bS by comparing interferometric data at millimeter and submillimeter wave bands with a simple model containing a single hot compact first-core-like component at the center with a large-scale cold dusty envelope. Our results provide an assessment of whether their physical properties agree with the theoretical expectations of the FHC stage or more evolved sources.

## 2. Observational Data

In this work, we compare our models to interferometric continuum observations of B1-bN and B1-bS obtained with the SMA and ALMA.

### 2.1. Observations with the SMA

#### 2.1.1. Observations with the SMA: 1.3 mm and 870 $\mu$m

The 1.3 mm and 870 $\mu$m observations centered at both B1-bN ($\alpha$(J2000) = $3^h33^m21\rlap{.}^s2$, $\delta$(J2000) = $31°07'43\rlap{.}''8$) and B1-bS ($\alpha$(J2000) = $3^h33^m21\rlap{.}^s4$, $\delta$(J2000) = $31°07'26\rlap{.}''4$) were simultaneously carried out on 2012 November 10 with the SMA in compact configuration (project code: 2012A-A006, PI: Hirano). However, one of the 1.3 mm receivers was malfunctioning at the time. The 1.3 mm observations used the 230 GHz receivers with six antennas and the 870 $\mu$m observations used the 400 GHz receivers with seven antennas. The spectral correlators covered a 2 GHz bandwidth in each of the two sidebands separated by 10 GHz. Each band is divided into 24 "chunks" of 104 MHz width. The visibility data were calibrated using the Millimeter Interferometer Reduction (MIR) software package. The absolute flux density scale was determined from observations of Uranus for the 230 GHz receivers and Neptune for the 400 GHz receivers. A pair of nearby compact radio sources, 3C84 (8.1 Jy at 1.3 mm and 6.9 Jy at 870 $\mu$m) and 3C111 (2.3 Jy at 1.3 mm and 2.0 at 870 $\mu$m), were used to calibrate the relative amplitude and phase. The bandpass of the data was calibrated by using Uranus. The calibrated visibility data were INVERTed and CLEANed using the MIRIAD package with natural weighting. This resulted in a synthesized beam of $3\rlap{.}''07 \times 2\rlap{.}''48$ for the 1.3 mm continuum map and $2\rlap{.}''04 \times 1\rlap{.}''69$ for the 870 $\mu$m map. The rms noise level of the continuum maps are 0.58 mJy beam$^{-1}$ and 2.60 mJy beam$^{-1}$, respectively.

#### 2.1.2. Observations with the SMA: 1.3 mm and 1.1 mm

Hirano & Liu (2014) presented 1.3 and 1.1 mm observations of B1-bN and B1-bS with the SMA in subcompact configuration. The phase-tracking center of the observations was $\alpha$(J2000) = $3^h33^m21\rlap{.}^s14$, $\delta$(J2000) = $31°07'35\rlap{.}''3$. We acquired the calibrated continuum visibility data sets and used INVERT and CLEAN in the MIRIAD package with natural weighting. The synthesized beams of the 1.3mm continuum map has the size of $5\rlap{.}''78 \times 3\rlap{.}''88$, and that of the 1.1 mm map has a size of $4\rlap{.}''82 \times 2\rlap{.}''47$. The rms noise level of the continuum maps are 1.0 and 3.05 mJy beam$^{-1}$, respectively.

In this paper, we present the results in continuum only. In order to increase the spatial dynamic ranges of the data, we also make a combined image of the 1.3 mm compact and subcompact configuration data. The final SMA continuum data are summarized in Table 1. Figure 1 shows the continuum emission from B1-bN and B1-bS observed with the SMA. The half-power widths of the sources are nearly the same as the beams. This implies that the sources are barely resolved with the current beams; the SMA images of B1-bN and B1-bS are close to point sources.

### 2.2. ALMA 870 $\mu$m (Band 7) Observations

We retrieved two ALMA 870 $\mu$m (Band 7) Cycle 3 observations with two configurations from the ALMA data

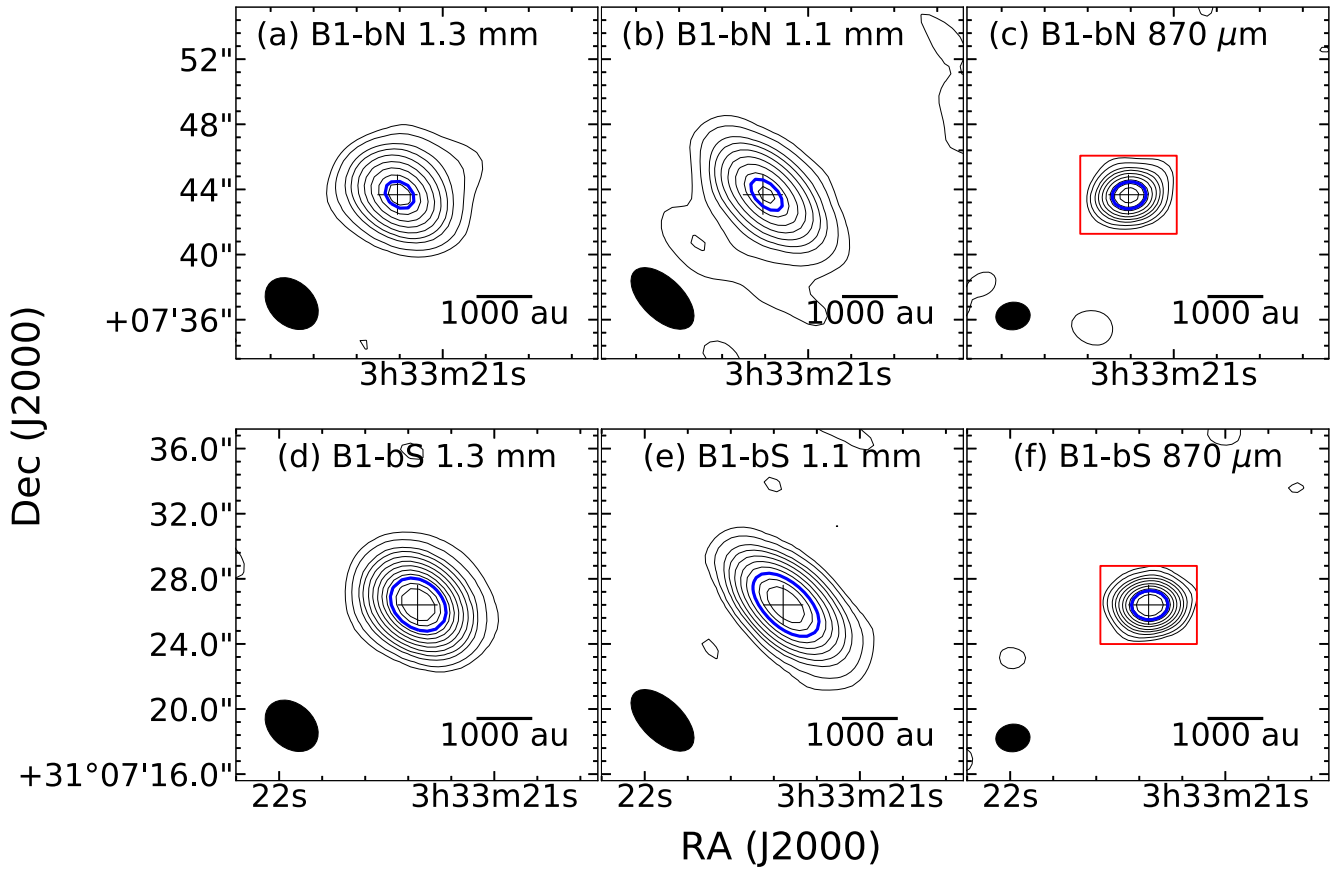

**Figure 1.** Maps of the continuum emission of B1-bN and B1-bS observed with the SMA. (a) The combined 1.3mm continuum map of B1-bN observed with the SMA. (b) The 1.1mm continuum map of B1-bN observed with the SMA at a subcompact configuration. (c) The 870 $\mu$m continuum map of B1-bN observed with the SMA at compact configuration. (d) The combined 1.3mm continuum map of B1-bS observed with the SMA. (e) The 1.1mm continuum map of B1-bS observed with the SMA at a subcompact configuration. (f) The 870 $\mu$m continuum map of B1-bS observed with the SMA at a compact configuration. The black contours of (a), (b), (d), and (e) start at $3\sigma$ and are drawn at $6\sigma$, $12\sigma$, $18\sigma$, $27\sigma$, $39\sigma$, $54\sigma$, $72\sigma$, $93\sigma$, $117\sigma$, $144\sigma$, and $217\sigma$. The black contours of (c) and (f) start at $3\sigma$ and are drawn at $12\sigma$, $27\sigma$, $54\sigma$, $93\sigma$, and $117\sigma$. The crosses denote the peak positions of B1-bN and B1-bS as derived from the two-dimensional Gaussian fitting. The black ellipses show the beam size of each configuration. The blue contours on each panel show the intensity at half of the peak. Red boxes on the 870 $\mu$m maps indicate the same field of view of the ALMA 870 $\mu$m compact (C40-4) configuration maps shown in Figure 2.

**Table 2**
The ALMA 870 $\mu$m Observations—Continuum

| Configuration | Beam Size | uv Range ($k\lambda$) | Flux Density (bN) (Jy) | Flux Density (bS) (Jy) | rms (mJy beam$^{-1}$) |
|---|---|---|---|---|---|
| C40-4 (compact)[a] | $0\rlap{.}''43 \times 0\rlap{.}''30$ | 13.9–753.7 | $3.23 \times 10^{-1}$ | $7.08 \times 10^{-1}$ | 0.50/0.80 |
| C36-7 (extended)[b] | $0\rlap{.}''054 \times 0\rlap{.}''037$ | 574.7–11494.3[c] | $3.21 \times 10^{-3}$ | $7.52 \times 10^{-3}$ | 0.20/0.20 |
| C40-4 + C36-7 | $0\rlap{.}''12 \times 0\rlap{.}''078$ | 13.9–11494.3 | $3.06 \times 10^{-1}$ | $6.58 \times 10^{-1}$ | 0.12/0.17 |

**Notes.**
[a] Project code: 2015.A.00012.S, PI: Gerin.
[b] Project code: 2015.1.00025.S, PI: Gerin.
[c] We exclude data with uv range <500 m.

archive centered on B1-bN and B1-bS. The first observation used the C36-7 configuration with baselines from 82 m up to 15 km (project code: 2015.1.00025.S, PI: Gerin), providing an angular resolution of $\sim 0\rlap{.}''04$. The second used the C40-4 configuration with baselines from 15 to 704 m (project code: 2015.A.00012.S, PI: Gerin), providing an angular resolution of $\sim 0\rlap{.}''4$. Both observations use the same pointing centers and spectral setups. See Gerin et al. (2017) for a more detailed description of the observations. We calibrated the data using the standard pipeline provided by the ALMA Regional Center (ARC) using CASA version 4.6.0. We follow the ARC suggestion in the imaging script and make maps excluding data with `uvrange` <500 m to reduce rms noise level of the continuum maps, since there are strong extended structures. Since the two data sets have vastly different uv coverages, we made the continuum images from the two data sets separately, as well as the combined images at higher spatial dynamic ranges. By imaging the combined data using Briggs weighting with robust parameters ranging from −2 to 2, we find that a robust parameter of 0 gives the best compromise between resolution and high spatial dynamic range. The properties of the ALMA continuum images are summarized in Table 2.

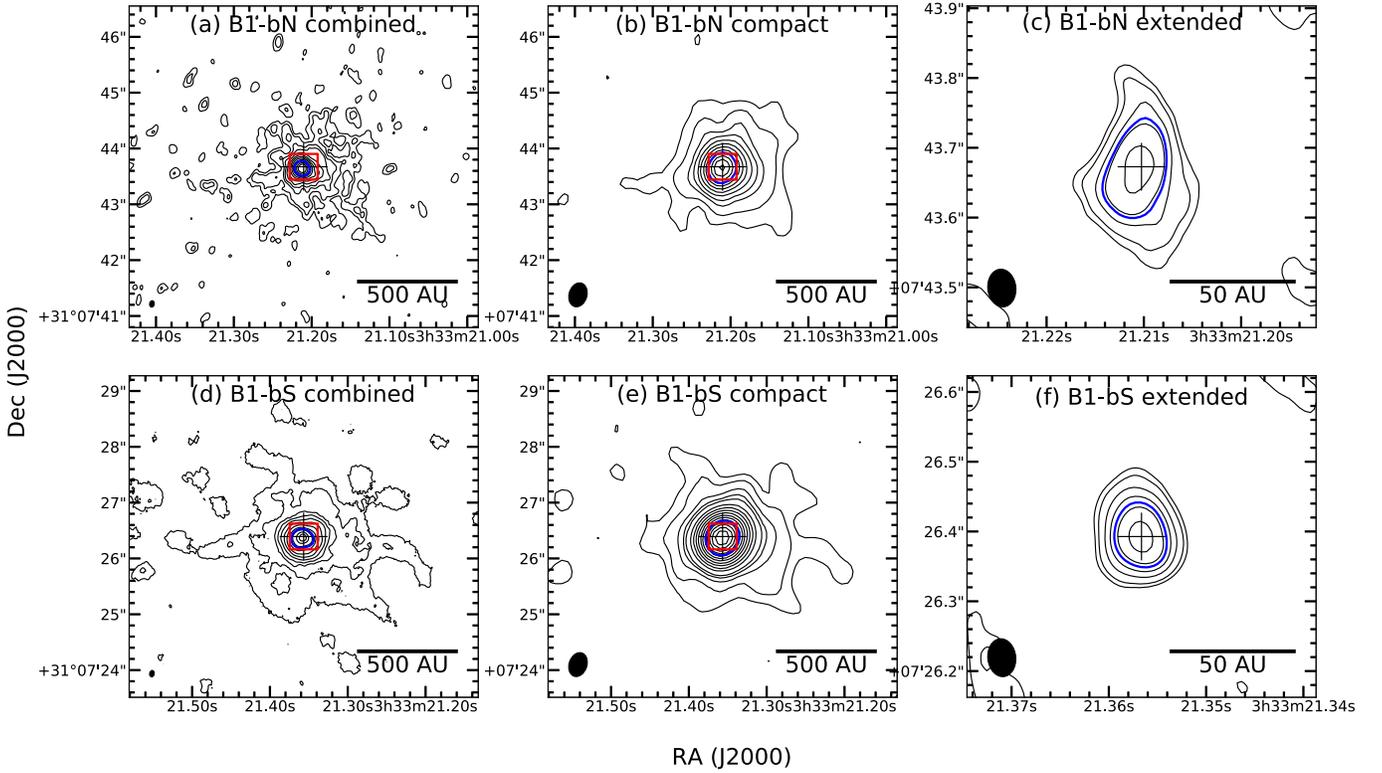

**Figure 2.** Maps of the continuum emission of B1-bN and B1-bS observed with ALMA at 870 $\mu$m. (a) The combined map of B1-bN, by combining the C40-4 (compact) and C36-7 (extended) configurations. (b) The map of B1-bN observed with C40-4 (compact) configuration. (c) The map of B1-bN observed with C36-7 (extended) configuration. (d) The combined map of B1-bS, by combining the C40-4 (compact) and C36-7 (extended) configurations. (e) The map of B1-bS observed with C40-4 (compact) configuration. (f) The map of B1-bS observed with C36-7 (extended) configuration. The black contours of (a), (b), (d), and (e) start at $3\sigma$ and are drawn at $6\sigma$, $12\sigma$, $27\sigma$, $54\sigma$, $93\sigma$, $144\sigma$, and $216\sigma$. The black contours of (c) and (f) start at $3\sigma$ and are drawn at $6\sigma$, $12\sigma$, $24\sigma$, $30\sigma$, and $42\sigma$. The crosses denote the peak positions of B1-bN and B1-bS. The black ellipses show the beam size of each configuration. The blue contours on each panel show the intensity at half of the peak. Red boxes on (a), (b), (d), and (e) indicate the same field of view of the C36-7 (extended) configuration maps.

Figure 2 shows the continuum emissions from B1-bN and B1-bS observed with ALMA. We refer to the ALMA C40-4 observation as the "ALMA compact observation" and the C36-7 observation as the "ALMA extended observation." The 870 $\mu$m flux density of the two sources observed with ALMA combined configuration are slightly lower than the values observed with the SMA; this is mainly due to the missing flux at large scales resolved by the longer baseline observations with ALMA. However, the spatial dynamic range of the ALMA data sets is much higher than those from the SMA.

## 3. Method and Modeling

In this paper, we test whether B1-bN and B1-bS can be described by a simple model of a central hot compact component surrounded by a dusty envelope. To compare the models with interferometric observations, we perform radiative transfer calculations and simulated observations to create synthesized images of our model (Section 3.1). The central hot compact component is set as a heating source, which is assumed to be a blackbody and has a fixed size of 4 au with different temperatures to represent different luminosities. We construct different physical structures of the envelope described in Sections 3.2 and 3.3. The temperature and central density range we tested in our model are in accordance with the theoretical predictions (Larson 1969; Masunaga et al. 1998). The density profile is assumed to be power-law-like similar to simulation results in Masunaga et al. (1998).

### 3.1. Procedure of Making Synthesized Images

Here, we describe the overall procedure to produce synthesized images of our models. We first construct dust-density profiles based on the physical structures of each model, and calculate their temperature profiles. We then generate intensity maps of our models using radiative transfer calculations. The visibilities of the intensity maps are generated and sampled based on real observations, with which synthesized images are produced. The five steps of creating a modeled synthesized image are shown in Figure 3.

#### 3.1.1. Step 1: Model Setup

We use the Monte Carlo–based code RADMC-3D (version 0.41; Dullemond et al. 2012a) for calculating temperature and running radiative transfer. To run RADMC-3D, we need to establish several inputs: a dust-density profile, a central heating source, a dust opacity, and an external radiation field. The dust-density profile represents the envelope part of our model and the central heating source represents the hot compact component at the center of the envelope (see detail in Sections 3.2 and 3.3). We adopt the $R_V = 5.5$ model from Weingartner & Draine (2001)[5] for the dust opacity $\kappa_\nu$, and assume it is independent of the radius within the envelope. The grain size distribution of the $R_V = 5.5$ model ranges from $\sim 3 \times 10^{-4}$ to

---
[5] The synthetic extinction curves developed by Weingartner & Draine (2001) are publicly available at https://www.astro.princeton.edu/~draine/dust/dustmix.html.

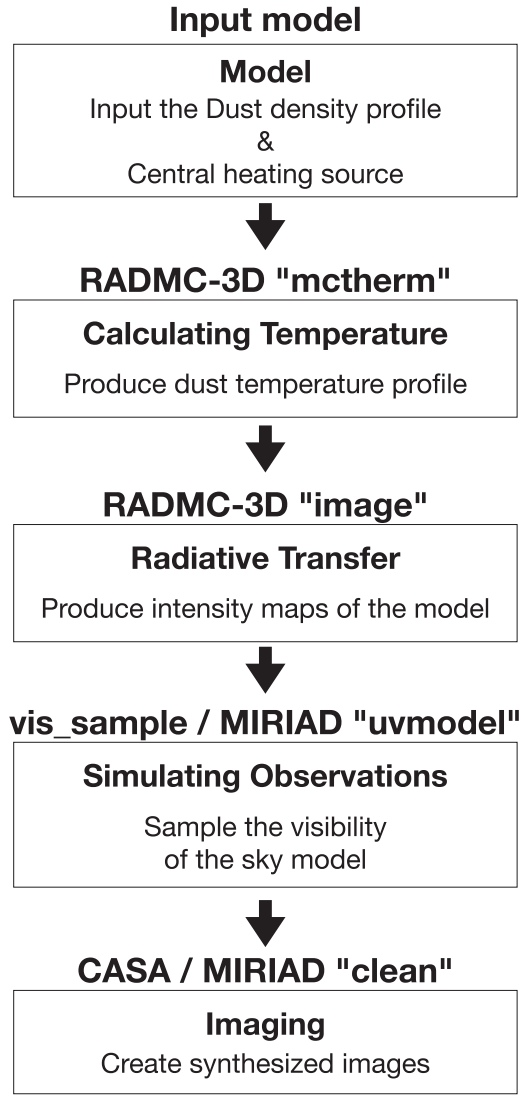

**Figure 3.** Steps for creating synthesized images of our first-core model.

∼2 μm (Figure 4 in Weingartner & Draine 2001), which simulates the grain size for dense core regions and is appropriate for this study since first cores are born in such dense molecular clouds. The dust opacity model provides $\kappa_\nu$, which covers the wavelength range recommended by RADMC-3D, $10^{-1}$ to $10^3$ μm, for a smooth Monte Carlo dust-temperature calculation. The interstellar radiation field around B1-bN and B1-bS is estimated to be 11.9 K from temperature maps of the Perseus molecular cloud (Zari et al. 2016), which we used for the external radiation field, $T_{\rm ext}$, in RADMC-3D.

*3.1.2. Step 2: Calculating Temperature*

The package `mctherm` in RADMC-3D is used to calculate the dust-temperature distributions of our model with the given density profile, size, and temperature of the central heating source, and the external radiation field. The central heating source is assumed to be a blackbody, i.e., no spectral model is given. If the luminosity of the central heating source is fixed, the temperature and size are degenerated. Therefore, we set the central heating source to a fixed size of 4 au with different temperatures to represent different luminosities. The wavelength coverage goes all the way from 0.1 to 1000 μm, and is logarithmically spaced. The package `mctherm` is in fact performing Monte Carlo simulations, which could give different results even with identical input parameters. We found that, when the number of photon packages is over $10^7$, the temperature profiles of all simulations are the same if the highest dust density of the envelope is less than $10^{-10}$ g cm$^{-3}$.

*3.1.3. Step 3: Radiative Transfer*

We used the `image` package in RADMC-3D to perform radiative transfer calculations on the input dust density and temperature profiles produced in Step 2 to generate theoretical sky-brightness images at the exact wavelengths of real observations. We set the number of photon packages to $10^7$.

*3.1.4. Step 4: Simulating Observations*

To simulate the ALMA observations, we use a Python package `vis_sample`[6] to sample visibilities from our theoretical sky-brightness images using the *uv* coverage of our observations. For SMA observations, we use the `uvmodel` task in MIRIAD. The noise is not included in the observing simulations.

*3.1.5. Step 5: Imaging*

The model visibilities can then be imaged using the same process and parameters as our observations. The imaging process is performed with the `tclean` task in CASA for the ALMA data and with the `invert` and `clean` tasks in MIRIAD for the SMA data.

*3.2. Model 1: Single Power-law Envelope*

We first try a single power-law envelope model with a hot compact central heating source of temperature $T_*$ and a fixed size of 4 au in diameter. The envelope shape is assumed to be spherical. No outflow cavity and envelope flattening are included. The dust density of the envelope at a radius $r$ is described by

$$\rho(r) = \rho_5 (r/5~{\rm au})^\alpha, \qquad (1)$$

where $\rho_5$ is the dust density at 5 au and $\alpha$ is the power-law index. The outer boundary of the envelope is fixed to a dust density, $\rho_{\rm out} = 10^{-24}$ g cm$^{-3}$, which is $n_{\rm H_2} \approx 6 \times 10^1$ cm$^{-3}$ using a gas-to-dust ratio of 100. The innermost cutoff radius of the density profiles is set to be the radius of the central heating source, which is a fixed value of 2 au in this study.

*3.3. Model 2: Broken Power-law Envelope with Outflow Cones*

In addition to the single power-law model, we also consider a model with outflow cones in the envelope, since the low density of the outflow cones may affect the radiative transfer result. Furthermore, a single power-law distribution of the envelope may produce an extremely large mass in the central region, which would not match observations; therefore, we also consider a broken power-law density profile with a shallower slope in the inner region.

Regarding the outflow cones, Gerin et al. (2015) have clearly detected molecular outflows of B1-bN and B1-bS in both $H_2CO$ and $CH_3OH$. Thus, to be more realistic, we construct a model with outflow cones where the dust density outside the outflow cones is described by Equation (1), while the density inside the outflow cones is the same as the density at the boundary of the envelope ($\rho_{out}$). Gerin et al. (2015) derived the opening angle of the outflow to be $23° \pm 8°$ for B1-bS, but did not provide measurements for B1-bN. Here, we adopt an opening angle of $\phi = 23°$ for B1-bS, and we use a smaller opening angle $\phi = 10°$ for B1-bN, which we measured directly from the map. We also adopt a position angle of the outflow cones to be $100°$ for both B1-bN and B1-bS, since the outflows appear to have a similar PA for both sources in the map. We ran some preliminary tests with different inclination angles, $i$, for the outflow cones of the two sources. We found that the model-intensity profiles are essentially the same for $i = 40°$–$80°$. Moreover, Gerin et al. (2015) show asymmetrical blue and redshifted outflow lobes for B1-bS and only a single blueshifted outflow lobe detected for B1-bN, implying that the inclination angles could not be $0°$ or $90°$ for both the sources. Thus, we simply adopt $i = 60°$ for both B1-bN and B1-bS. In summary, considering outflow cones in the envelope, we add fixed bipolar outflow cones with $i = 60°$, $\phi = 10°$, and $PA = 100°$ for B1-bN, and $i = 60°$, $\phi = 23°$, and $PA = 100°$ for B1-bS.

---
[6] The vis_sample Python package is provided by Ian Czekala and is publicly available at https://github.com/AstroChem/vis_sample.

Regarding the power-law index of the inner envelope, we adopt $\alpha_0 = -1.5$ from the inside-out collapse model (Shu 1977). When inner-envelope flattening is included together with outflow cones, the density of the envelope at a radius, $r$, outside the outflow cones is set to

$$\rho(r) = \rho_{b,5}(R_b/5 \text{ au})^\alpha (r/R_b)^{-1.5} \quad \text{if} \quad r \leqslant R_b$$
$$\rho(r) = \rho_{b,5}(r/5 \text{ au})^\alpha \quad \text{if} \quad r > R_b, \quad (2)$$

where $\rho_{b,5}$ is the density at 5 au adjusted for the broken power law, and $R_b$ is the breaking radius of the broken power law. Since we fix the size of the central source to 4 au, the central density in our model, $\rho(2au)$, is a function of both $\rho_{b,5}$ and $R_b$.

Figure 4 shows an overall structure of our model, and Table 3 lists all the parameters used for each model.

## 4. Results and Analyses

### 4.1. The Initial Parameter Search Ranges and Grids

In order to find appropriate parameter values for our model, we run models with ranges of different parameter values and compare them with observations. We begin by testing typical fiducial values predicted from FHC theories. We then quantify how well the models reproduce the observations by calculating the $\chi^2$ value as follows. Since the SMA images of B1-bN and B1-bS are close to point sources (see Section 2.1), we adopt $\chi^2 = ((\text{flux density}_{obs}-\text{flux density}_{model})/\text{error})^2$ for the comparison. The error is the error estimated from two-dimensional Gaussian fittings to the sources (Table 1). We define $\chi^2_{SMA} = \chi^2_{1.3\,mm} + \chi^2_{1.1\,mm} + \chi^2_{870\mu m}$, of which the subscripts indicate the wavelengths of our SMA observations. For the ALMA observations of a given model, we calculate the $\chi^2_{ALMA}$ values by comparing the annually averaged intensity profiles of the model with that of the combined ALMA observations, where the intensity is larger than $3\sigma$. Ideally, a model with both a small $\chi^2_{SMA}$ and $\chi^2_{ALMA}$ means it matches well to the observed data.

We plug in the fiducial values of $\alpha = -2.0$ (adopted from Masunaga et al. 1998 and the Larson–Penston solution (Larson 1969; Penston 1969)), $\rho_5 = 10^{-11}$ g cm$^{-3}$ and $T_* = 500$ K (Figure 1 in Inutsuka 2012) into Model 1. We find these parameters do not provide a good match to the observations (with $\chi^2_{ALMA} > 150$ and $\chi^2_{SMA} > 10$). Plugging in the same fiducial values into Model 2 (with the consideration of the outflow cones), the resulting $\chi^2$ values are also very large for both b1-bN and b1-bS: $\chi^2_{SMA} = 9.34$ and $\chi^2_{ALMA} = 72.49$ for b1-bN, and $\chi^2_{SMA} = 8.84$ and $\chi^2_{ALMA} = 88.74$ for b1-bS (see the profiles in Figure 7). Therefore, we perform a search using both Model 1 and Model 2 by varying each parameter with a large range and spacing in a systematical way to find the best model (Table 4). We find that the resulting $\chi^2$ values for Model 1 are still quite large, with no $\chi^2_{SMA} < 2$ and no $\chi^2_{ALMA} < 10$, while for Model 2, $\chi^2_{SMA}$ and $\chi^2_{ALMA}$ are mostly in the range of 0.1–5 and 0.1–30, respectively. Overall, Model 2 has parameter ranges with smaller $\chi^2_{SMA}$ and $\chi^2_{ALMA}$. Our results from this wider search indicate that Model 1 may not be appropriate to describe the source within the given parameter range. Hence, we only discuss the results from Model 2, i.e., considering a broken power-law envelope with outflow cones.

### 4.2. The Narrowed Down Parameter Ranges and Grids

The most appropriate parameters for B1-bN and B1-bS are found with the lowest $\chi^2_{SMA}$ and $\chi^2_{ALMA}$ in Table 4 from the aforementioned wide-range parameter search. We narrow down the parameter space for B1-bN and B1-bS separately (Tables 5 and 6) using the models with $\chi^2_{SMA} < 0.3$ and $\chi^2_{ALMA} < 5$, which are so-called "good models." In addition, the ALMA extended configuration data are the most important data. Since the angular resolution of the ALMA extended configuration data ($\sim 0''.045$) is much higher than that of the combined data ($\sim 0''.1$), any model that does not reproduce the ALMA extended data is not meaningful. To further investigate whether these "good models" are consistent with the observations, we calculate $\chi^2_{extended}$ from the intensity profile of the ALMA extended (C36-7) configuration data independently. This $\chi^2_{extended}$ is calculated from the difference between the annually averaged intensity profiles of the model and ALMA extended observations. Tables 7 and 8 list the resulting values of these models (with $\chi^2_{extended} < 150$) for B1-bN and B1-bS, respectively. We show that many models with small $\chi^2_{SMA}$ and $\chi^2_{ALMA}$ have $\chi^2_{extended} \gg 10$ because of their significant difference in $uv$ range. Therefore, considering the ALMA extended configuration observation independently is needed to pin down the parameters for the best model. We found that a small difference between the model and observation at the center could have a large contribution in $\chi^2_{extended}$. To obtain the best profile matching the observation at all radii, we visually inspect the annually averaged profiles with $\chi^2_{extended} < 10$ from the "good models" and find the best model for B1-bN and B1-bS. In Table 9, we summarize the $\chi^2$ values and parameters of the best models. We note that the best model for B1-bS is the one with the smallest $\chi^2_{extended}$, while that for B1-bN is the one with the smallest $\chi^2_{ALMA}$.

### 4.3. The Best Model

Figures 5 and 6 show the ALMA 870 $\mu$m intensity maps of observations, the corresponding best models, and the residual maps for both sources. The overall residuals are <10%, and the maximum residuals are <20% of the observed peak intensity.

Table 3
The Parameters of Our FHC Models

| Parameter | | Values | Envelope Model |
|---|---|---|---|
| Central heating source: | | | |
| Size | $D_*$ (au) | 4 | |
| Temperature | $T_*$ (K) | 100–2000 | |
| Envelope: | | | |
| Density at 5 au | $\rho_5$ (g cm$^{-3}$) | $10^{-12} \sim 10^{-10}$ | Model 1 |
| Density at 5 au adjusted for the broken power law | $\rho_{b,5}$ (g cm$^{-3}$) | $10^{-12} \sim 10^{-10}$ | Model 2 |
| Power-law index | $\alpha$ | $-1.5 \sim -3.5$ | Model 1 and Model 2 |
| Outflow cones' opening angle | $\phi$ (°) | 23 (bS), 10 (bN) | Model 2 |
| Outflow cones' inclination | $i$ (°) | 60 | Model 2 |
| Outflow cones' position angle | PA (°) | 10 | Model 2 |
| Breaking radius | $R_b$ (au) | $0 \sim 150$ | Model 2 |
| Power-law index at $r < R_b$ | $\alpha_0$ | $-1.5$ | Model 2 |
| Dust and interstellar medium properties: | | | |
| Absorption cross section ($R_v = 5.5$) | $\kappa_\nu$ (cm$^2$ g$^{-1}$) | 0.0039 at 870 $\mu$m | |
| (Per unit gas + dust mass) | | 0.0028 at 1.1 mm | |
| | | 0.002 at 1.3 mm | |
| Density at the boundary | $\rho_{out}$ (g cm$^{-3}$) | $10^{-24}$ | |
| External radiation field | $T_{ext}$ (K) | 11.9 | |

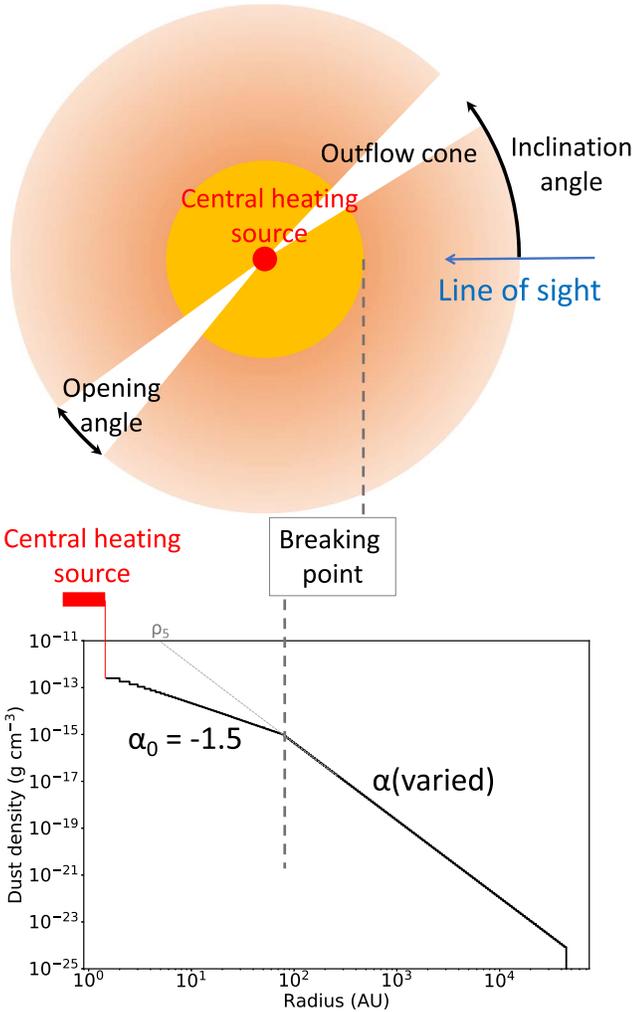

Figure 4. The overall structure of our model.

The residuals show some symmetric/asymmetric features, which may be due to nonspherical morphology that our simple spherical model cannot reproduce. We could further reduce the residuals by refining the outflows position angle. We find that the rms is minimized for a PA of 105° for both B1-bN and B1-bS, as shown by the model and residual images presented in Figures 5 and 6. Note that we fixed the PA = 100° when we selected the best models. However, the accuracy of the outflow cones position angle does not affect the fitting results, since we calculate $\chi^2$ by comparing the annually averaged profiles.

Figure 7 shows the annually averaged profiles of ALMA data together with the models. The best model for B1-bN well reproduces the observed intensity profiles of both the combined and extended configuration. For B1-bS, the best model overall reproduces the combined intensity profile, while it tends to be higher than the observed intensity at $r \sim 20$–30 au, and lower at $r > 40$ au. For the purpose of comparison, we also plot the profiles of Model 2 with the fiducial parameters. The intensity in the central region is low because the central hot compact source is very embedded; the envelope central density $\rho_{(2\ au)}$ for the fiducial model is much higher than our best models. A local minimum at $r \sim 80$ au is found in the fiducial profile. This is originated from radiative transfer. We find that the slope in the temperature profile at $r \sim 50$–100 au is steeper than that of the density profile. The characteristics of the fiducial model above indicate that the envelope central density, power-law index, and the breaking radius need to be evaluated. Figure 8 shows the best model dust-density profiles and dust-temperature profiles of the envelope for B1-bN and B1-bS. The central density $\rho_{(2\ au)}$ of the best models are $4.29 \times 10^{-13}$ and $3.38 \times 10^{-13}$ g cm$^{-3}$ for B1-bN and B1-bS, respectively. Discussions of these parameters and profiles are reported in Section 5.4.

Figure 9 shows the visibilities of observations and our best model. The error bar in the figures represents the rms of the visibilities in each bin. The observational noise is the 1-baseline noise with the ALMA continuum bandwidth. For

Table 4
Range and Grid of the Coarse Parameter Search for Both the B1-bN and B1-bS Models

| $\alpha$ | −1.5 | −1.8 | −2.0 | −2.3 | −2.5 | −2.8 | −3.0 | −3.3 | −3.5 |
|---|---|---|---|---|---|---|---|---|---|
| $\rho_{b,5}$ (g cm$^{-3}$) | $10^{-12}$ | $10^{-11.8}$ | $10^{-11.5}$ | $10^{-11.3}$ | $10^{-11}$ | $10^{-10.8}$ | $10^{-10.5}$ | $10^{-10.3}$ | $10^{-10}$ |
| $T_*$ (K) | 100 | 300 | 500 | 700 | 1000 | 1500 | 2000 | | |
| $R_b$ (au) | 0 | 50 | 100 | 200 | | | | | |

Table 5
Range and Grid of the Refined Parameter Search for the B1-bN Model

| $\alpha$ | −2.7 | −2.8 | −2.9 | −3.0 | −3.1 | | | |
|---|---|---|---|---|---|---|---|---|
| $\rho_{b,5}$ (g cm$^{-3}$) | $10^{-12}$ | $10^{-11.9}$ | $10^{-11.8}$ | $10^{-11.7}$ | $10^{-11.6}$ | | | |
| $T_*$ (K) | 300 | 350 | 450 | 500 | 550 | 600 | 650 | 700 |
| $R_b$ (au) | 0 | 10 | 20 | 30 | 40 | 50 | 60 | |

**Note.** Parameter values encircled with open rectangles denote the best parameter set obtained from the coarse parameter search tabulated in Table 4.

Table 6
Range and Grid of the Refined Parameter Search for the B1-bS Model

| $\alpha$ | −3.0 | −3.1 | −3.2 | −3.3 | −3.4 | | | |
|---|---|---|---|---|---|---|---|---|
| $\rho_{b,5}$ (g cm$^{-3}$) | $10^{-11.3}$ | $10^{-11.2}$ | $10^{-11.1}$ | $10^{-11.0}$ | $10^{-10.9}$ | | | |
| $T_*$ (K) | 300 | 350 | 450 | 500 | 550 | 600 | 650 | 700 |
| $R_b$ (au) | 50 | 60 | 70 | 80 | 90 | 100 | | |

**Note.** Parameter values encircled with open rectangles denote the best parameter set obtained from the coarse parameter search tabulated in Table 4. In this table, $R_b = 50$ and 100 au both produce the lowest $\chi^2_{SMA}$ and $\chi^2_{ALMA}$.

the SMA data, due to both B1-bN and B1-bS being within the field of view in each pointing, we subtract the observed clean components of one of the sources from the observed visibilities with the MIRIAD task UVMODEL. These results show that the flux calibration between the SMA and ALMA data are in agreement with each other. Overall, our models are consistent with the observations for $uv$ distances <1000 k$\lambda$. For B1-bS, the model visibilities at 500 k$\lambda$ < $uv$ < 800 k$\lambda$ are in between the compact and extended configuration data. The amplitude of B1-bS in the compact configuration drops significantly at a $uv$ distance of 350 k$\lambda$, which introduces the amplitude inconsistency (within error) between the compact and extended data. We decide not to refine the gains since the B1-bN data were observed with B1-bS simultaneously; if we adjust the gains to match the visibilities of the two configurations for B1-bS, the amplitude difference will show up in the B1-bN visibilities. The observed visibilities show constant values at a $uv$ distance of >1000 k$\lambda$ for B1-bN and >800 k$\lambda$ for B1-bS. This implies the presence of a point source in both B1-bN and B1-bS with a flux density of 0.02 Jy. The model-visibility profiles show a sinc-like profile at $uv$ distances >1000 k$\lambda$. This is an artificial effect caused by our model setting used for the radiative transfer calculation. Since the resolution of the sky-brightness model is set to be 1 au, the central source with a diameter of 4 au spreads across a few pixels in the sky map. When the sky map is Fourier transformed, the central source with a finite diameter produces the "artificial" sinc-like profile in the visibilities. We examined the sky map and found that the enclosed intensity with $R$ <2 au is ∼0.02 Jy, consistent with the amplitude of observed visibilities.

The flux densities from the sky brightness of the best models and photometric data of B1-bN and B1-bS are shown in Figure 10. The model flux density values are derived within the same beam sizes of the Herschel PACS, CSO SHARC 350 $\mu$m, and JCMT SCUBA 850 $\mu$m observations. Our best models can overall reproduce the 100–1000 $\mu$m flux density. The observations at 350 and 850 $\mu$m with the CSO SHARC and JCMT SCUBA, respectively, are slightly higher than the best models. This may be due to diffuse background contamination from ambient emission in the Perseus molecular cloud, which is not reproduced in our model. The 350 and 850 $\mu$m data points were measured using an aperture size of 20″ × 20″, which is much larger than the source size (∼1″; Hirano & Liu 2014). In Figure 10, we also show that, for the SMA 870 $\mu$m, the combined ALMA 870 $\mu$m, the ALMA 870 $\mu$m with extended configuration, and the SMA 1.1 and 1.3 mm, the flux densities of the two sources derived from the best simulated synthesized images are consistent with the observations. The low amplitude at 870 $\mu$m with the ALMA extended configuration is due to the effect of the missing flux, which is also reproduced with simulation.

## 5. Discussion

### 5.1. The Presence of the Central Point Sources

The presence of a central point source is crucial for determining the properties of an FHC. In previous studies, Pezzuto et al. (2012) and Hirano & Liu (2014) fit the SEDs of B1-bN and B1-bS with one or two isothermal graybodies without characterizing the embedded central source, while our work used radiative transfer modeling of an explicit hot compact component surrounded by a dusty envelope. Our best models show that, with a fixed size of 4 au, the temperature of the hot compact component in the two first hydrostatic core candidates is 450 K for B1-bN and 550 K for B1-bS. As we showed in Section 4.3, such compact components would appear as ∼0.02 Jy point sources, which is consistent with the observed visibilities (Figure 9). Therefore, our radiative transfer modeling has successfully characterized the properties of the embedded sources. It is desirable to examine other first hydrostatic core candidates to explore the existence of first cores and their physical properties.

Table 7
The $\chi^2$ of the "Good Models" for B1-bN

| Parameters: | | | | | $\chi^2$ Values: | | | |
|---|---|---|---|---|---|---|---|---|
| $\alpha$ | $\rho_{b,5}$ (g cm$^{-3}$) | $T_*$ (K) | $R_b$ (au) | $\rho$(2 au) (g cm$^{-3}$) | $\chi^2_{\rm SMA}$ | $\chi^2_{\rm ALMA}$ | $\chi^2_{\rm extended}$ | |
| −2.8 | $10^{-11.9}$ | 600 | 10 | $2.02 \times 10^{-12}$ | 0.14 | 0.82 | 2.20 | |
| −2.9 | $10^{-11.8}$ | 700 | 0 | $2.26 \times 10^{-11}$ | 0.23 | 2.81 | 2.60 | |
| −2.9 | $10^{-11.7}$ | 450 | 40 | $4.29 \times 10^{-13}$ | 0.29 | 0.05 | 3.11 | The best model |
| −2.8 | $10^{-11.9}$ | 550 | 10 | $2.02 \times 10^{-12}$ | 0.12 | 0.16 | 4.81 | |
| −2.7 | $10^{-12}$ | 400 | 30 | $4.60 \times 10^{-13}$ | 0.25 | 1.00 | 5.34 | |
| −2.8 | $10^{-11.8}$ | 350 | 50 | $3.14 \times 10^{-13}$ | 0.18 | 3.47 | 7.11 | |
| −2.9 | $10^{-11.8}$ | 650 | 0 | $2.26 \times 10^{-11}$ | 0.15 | 1.59 | 7.15 | |
| −3 | $10^{-11.7}$ | 700 | 0 | $3.12 \times 10^{-11}$ | 0.18 | 4.00 | 7.46 | |
| −2.8 | $10^{-11.9}$ | 600 | 0 | $4.98 \times 10^{-12}$ | 0.12 | 0.34 | 7.57 | |
| −2.9 | $10^{-11.8}$ | 650 | 10 | $2.37 \times 10^{-12}$ | 0.14 | 2.50 | 7.82 | |
| −3 | $10^{-11.6}$ | 550 | 30 | $6.76 \times 10^{-13}$ | 0.21 | 2.29 | 11.96 | |
| −3 | $10^{-11.7}$ | 700 | 10 | $2.79 \times 10^{-12}$ | 0.16 | 4.57 | 12.64 | |
| −2.8 | $10^{-11.8}$ | 350 | 40 | $4.20 \times 10^{-13}$ | 0.16 | 4.76 | 15.84 | |
| −2.8 | $10^{-11.9}$ | 550 | 20 | $8.21 \times 10^{-13}$ | 0.16 | 0.90 | 16.54 | |
| −2.8 | $10^{-11.8}$ | 450 | 30 | $6.10 \times 10^{-13}$ | 0.15 | 0.68 | 17.50 | |
| −2.9 | $10^{-11.6}$ | 300 | 60 | $3.06 \times 10^{-13}$ | 0.24 | 2.17 | 18.58 | |
| −2.8 | $10^{-11.8}$ | 350 | 60 | $2.48 \times 10^{-13}$ | 0.13 | 0.87 | 30.28 | |
| −2.7 | $10^{-12}$ | 350 | 50 | $2.49 \times 10^{-13}$ | 0.21 | 1.94 | 40.78 | |
| −2.9 | $10^{-11.7}$ | 450 | 30 | $6.42 \times 10^{-13}$ | 0.14 | 0.37 | 42.21 | |
| −2.7 | $10^{-12}$ | 400 | 40 | $3.26 \times 10^{-13}$ | 0.25 | 0.59 | 44.57 | |
| −2.7 | $10^{-12}$ | 350 | 60 | $2.00 \times 10^{-13}$ | 0.29 | 1.37 | 53.87 | |
| −2.8 | $10^{-11.8}$ | 400 | 30 | $6.10 \times 10^{-13}$ | 0.14 | 2.33 | 57.79 | |
| −2.9 | $10^{-11.8}$ | 600 | 20 | $9.00 \times 10^{-13}$ | 0.12 | 2.80 | 58.32 | |
| −2.7 | $10^{-12}$ | 500 | 0 | $1.19 \times 10^{-11}$ | 0.16 | 3.74 | 59.42 | |
| −2.9 | $10^{-11.7}$ | 400 | 60 | $2.43 \times 10^{-13}$ | 0.11 | 0.10 | 62.37 | |
| −2.8 | $10^{-11.8}$ | 400 | 60 | $2.48 \times 10^{-13}$ | 0.26 | 0.40 | 64.15 | |
| −2.7 | $10^{-12}$ | 450 | 10 | $1.72 \times 10^{-12}$ | 0.18 | 3.44 | 65.40 | |
| −2.9 | $10^{-11.7}$ | 450 | 50 | $3.14 \times 10^{-13}$ | 0.19 | 0.24 | 71.21 | |
| −2.8 | $10^{-11.9}$ | 450 | 40 | $3.33 \times 10^{-13}$ | 0.19 | 0.32 | 71.62 | |
| −3 | $10^{-11.6}$ | 600 | 20 | $1.24 \times 10^{-12}$ | 0.26 | 1.98 | 72.64 | |
| −2.7 | $10^{-12}$ | 400 | 60 | $2.00 \times 10^{-13}$ | 0.13 | 0.74 | 84.67 | |
| −2.9 | $10^{-11.7}$ | 450 | 60 | $2.43 \times 10^{-13}$ | 0.17 | 0.27 | 103.24 | |
| −2.8 | $10^{-11.9}$ | 450 | 50 | $2.49 \times 10^{-13}$ | 0.26 | 0.81 | 103.51 | |
| −2.7 | $10^{-12}$ | 450 | 60 | $2.00 \times 10^{-13}$ | 0.25 | 0.54 | 107.76 | |
| −2.9 | $10^{-11.7}$ | 500 | 20 | $1.13 \times 10^{-12}$ | 0.11 | 0.28 | 111.73 | |
| −2.8 | $10^{-11.9}$ | 450 | 60 | $1.97 \times 10^{-13}$ | 0.28 | 0.26 | 113.31 | |
| −3 | $10^{-11.6}$ | 650 | 10 | $3.51 \times 10^{-12}$ | 0.18 | 1.39 | 114.86 | |
| −3 | $10^{-11.6}$ | 500 | 50 | $3.14 \times 10^{-13}$ | 0.20 | 1.05 | 117.16 | |
| −2.9 | $10^{-11.7}$ | 600 | 10 | $2.99 \times 10^{-12}$ | 0.23 | 0.28 | 121.88 | |
| −2.8 | $10^{-11.9}$ | 500 | 50 | $2.50 \times 10^{-13}$ | 0.27 | 0.96 | 149.45 | |

### 5.2. Luminosity of B1-bN and B1-bS

#### 5.2.1. The Intrinsic Luminosity of the FHC

We calculate the luminosities of the hot compact sources via the Stefan–Boltzmann law: $L_* = \sigma T_*^4 \, 4\pi \left(\frac{1}{2} D_*\right)^2$. We find the values derived from the best models to be 6.70 $L_\odot$ for B1-bN and 14.97 $L_\odot$ for B1-bS (Table 10). The key parameter of the central hot compact source in our radiative transfer calculation is the luminosity, i.e., the energy supplied by the central source. The derived luminosity from our best model can be treated as the intrinsic luminosity of the FHC. The temperature and size are consistent with theoretical expectations of an FHC, which have a temperature lower than 2000 K and a size of ∼5 au (Masunaga et al. 1998), suggesting that B1-bN and B1-bS are in an early evolutionary stage. Since the size of the central compact source is fixed in the model, the luminosity difference between the two best models for B1-bN and B1-bS originates from the temperature difference. However, the FHC luminosity may be variable due to episodic accretion, which generates shocks and heats up the FHC surface in a short timescale, thus making it an unreliable evolutionary indicator.

#### 5.2.2. The Bolometric Luminosity of the Source

Since an FHC is very embedded, the radiation emitted from the FHC surface can be completely absorbed in the very opaque material immediately outside the FHC (Larson 1969). The bolometric luminosities, $L_{\rm bol}$, of our best models are calculated to be 0.18 and 0.26 $L_\odot$ for B1-bN and B1-bS, respectively (Table 10), by integrating the sky-brightness flux density within an aperture of 13″ from 10 to 10,000 $\mu$m. These $L_{\rm bol}$ values are consistent with the observed bolometric luminosity range of first-core candidates, which is 0.1–0.25 $L_\odot$. The $L_{\rm bol}$ of the two sources reported by Hirano & Liu (2014) are 0.15 $L_\odot$ for B1-bN and 0.31 $L_\odot$ for B1-bS. However, the adopted distance of the sources in Hirano & Liu (2014) is 230 pc (Černis & Straižys 2003), while we use 301 pc (Zucker et al. 2018). Scaling to 301 pc, the $L_{\rm bol}$ of the two sources reported by Hirano & Liu (2014) are scaled to 0.25 $L_\odot$ for B1-bN and 0.53 $L_\odot$ for B1-bS. The luminosity difference between Hirano & Liu (2014) and our work mainly comes from the difference in the 70 $\mu$m flux density. The 70 $\mu$m flux density observed with Herschel PACS (it is the upper limit for B1-bN) is higher than the model flux density at the same wavelength. In the case of B1-bN, the $L_{\mathrm{bol}}$ could be overestimated by Hirano & Liu (2014) because the graybody curve goes through the upper limit point. Taking this into account, there is no significant difference between observed and model $L_{\mathrm{bol}}$ values. On the other hand, in the case of B1-bS, the model flux density is 1 order of magnitude lower than the observed flux density detected by Herschel PACS at 70 $\mu$m. The 70 $\mu$m emission could be affected by dust scattering, depending on the dust grain size distribution (Ysard et al. 2019); thus, the luminosity may not be well constrained. Since the flux density at each wavelength is independently modeled, our best model parameters could still overall reproduce the 100–1000 $\mu$m flux density, and properly described the SMA and ALMA observations.

Table 8
The $\chi^2$ of the "Good Models" for B1-bS

| Parameters: | | | | | $\chi^2$ Values: | | | |
|---|---|---|---|---|---|---|---|---|
| $\alpha$ | $\rho_{b,5}$ (g cm$^{-3}$) | $T_*$ (K) | $R_b$ (au) | $\rho(2\mathrm{au})$ (g cm$^{-3}$) | $\chi^2_{\mathrm{SMA}}$ | $\chi^2_{\mathrm{ALMA}}$ | $\chi^2_{\mathrm{extended}}$ | |
| **−3.3** | **$10^{-10.9}$** | **550** | **80** | **$3.38 \times 10^{-13}$** | **0.20** | **2.25** | **4.29** | The best model |
| −3.1 | $10^{-11.3}$ | 550 | 60 | $3.72 \times 10^{-13}$ | 0.23 | 3.69 | 6.80 | |
| −3.2 | $10^{-11.1}$ | 550 | 70 | $3.54 \times 10^{-13}$ | 0.21 | 3.15 | 7.92 | |
| −3 | $10^{-11.3}$ | 500 | 90 | $2.59 \times 10^{-13}$ | 0.27 | 2.56 | 9.24 | |
| −3.2 | $10^{-11.1}$ | 550 | 80 | $2.82 \times 10^{-13}$ | 0.20 | 3.53 | 10.51 | |
| −3.2 | $10^{-11}$ | 550 | 90 | $2.90 \times 10^{-13}$ | 0.29 | 2.24 | 10.54 | |
| −3 | $10^{-11.3}$ | 450 | 100 | $2.21 \times 10^{-13}$ | 0.24 | 4.38 | 13.10 | |
| −3.2 | $10^{-11}$ | 500 | 100 | $2.43 \times 10^{-13}$ | 0.21 | 2.71 | 17.29 | |
| −3.1 | $10^{-11.2}$ | 550 | 70 | $3.66 \times 10^{-13}$ | 0.28 | 1.97 | 18.17 | |
| −3.1 | $10^{-11.2}$ | 500 | 90 | $2.45 \times 10^{-13}$ | 0.24 | 3.48 | 18.88 | |
| −3.3 | $10^{-10.9}$ | 550 | 90 | $2.74 \times 10^{-13}$ | 0.19 | 2.84 | 19.74 | |
| −3.2 | $10^{-11.1}$ | 600 | 80 | $2.82 \times 10^{-13}$ | 0.23 | 4.56 | 23.33 | |
| −3 | $10^{-11.3}$ | 500 | 100 | $2.21 \times 10^{-13}$ | 0.24 | 3.00 | 30.12 | |
| −3.2 | $10^{-11}$ | 500 | 80 | $3.55 \times 10^{-13}$ | 0.23 | 2.38 | 32.31 | |
| −3.2 | $10^{-11}$ | 550 | 100 | $2.43 \times 10^{-13}$ | 0.24 | 2.62 | 32.91 | |
| −3.3 | $10^{-10.9}$ | 600 | 90 | $2.74 \times 10^{-13}$ | 0.27 | 4.11 | 34.38 | |
| −3.1 | $10^{-11.2}$ | 550 | 90 | $2.45 \times 10^{-13}$ | 0.23 | 3.02 | 35.80 | |
| −3.1 | $10^{-11.2}$ | 500 | 70 | $3.66 \times 10^{-13}$ | 0.24 | 3.05 | 37.75 | |
| −3.2 | $10^{-11.1}$ | 550 | 90 | $2.31 \times 10^{-13}$ | 0.22 | 4.48 | 43.07 | |
| −3.3 | $10^{-10.9}$ | 550 | 100 | $2.26 \times 10^{-13}$ | 0.19 | 4.02 | 45.62 | |
| −3.4 | $10^{-10.9}$ | 600 | 60 | $4.43 \times 10^{-13}$ | 0.22 | 4.44 | 45.79 | |
| −3.1 | $10^{-11.2}$ | 550 | 100 | $2.07 \times 10^{-13}$ | 0.21 | 3.50 | 54.17 | |
| −3 | $10^{-11.3}$ | 500 | 70 | $3.78 \times 10^{-13}$ | 0.29 | 2.66 | 54.17 | |
| −3.2 | $10^{-11.1}$ | 600 | 60 | $4.60 \times 10^{-13}$ | 0.27 | 3.93 | 76.04 | |
| −3.2 | $10^{-11.1}$ | 600 | 100 | $1.93 \times 10^{-13}$ | 0.18 | 4.16 | 77.19 | |
| −3.2 | $10^{-11.2}$ | 600 | 50 | $4.98 \times 10^{-13}$ | 0.21 | 4.86 | 78.76 | |
| −3.1 | $10^{-11.3}$ | 600 | 90 | $1.94 \times 10^{-13}$ | 0.19 | 4.76 | 82.05 | |
| −3.1 | $10^{-11.3}$ | 600 | 50 | $4.98 \times 10^{-13}$ | 0.22 | 3.42 | 83.69 | |
| −3.3 | $10^{-10.9}$ | 550 | 70 | $4.30 \times 10^{-13}$ | 0.22 | 1.99 | 89.60 | |
| −3.2 | $10^{-11.1}$ | 550 | 60 | $4.60 \times 10^{-13}$ | 0.22 | 2.83 | 102.78 | |
| −3.3 | $10^{-10.9}$ | 500 | 70 | $4.30 \times 10^{-13}$ | 0.25 | 3.08 | 112.85 | |
| −3.1 | $10^{-11.3}$ | 550 | 50 | $4.98 \times 10^{-13}$ | 0.25 | 3.24 | 113.89 | |
| −3.1 | $10^{-11.2}$ | 550 | 60 | $4.68 \times 10^{-13}$ | 0.28 | 1.79 | 114.37 | |
| −3.1 | $10^{-11.2}$ | 500 | 60 | $4.68 \times 10^{-13}$ | 0.24 | 3.59 | 145.41 | |
| −3.2 | $10^{-11}$ | 500 | 70 | $4.45 \times 10^{-13}$ | 0.23 | 3.68 | 147.36 | |

Table 9
Parameters and $\chi^2$ of the Best Models

| | $\chi^2_{\mathrm{SMA}}$ | $\chi^2_{\mathrm{ALMA}}$ | $\chi^2_{\mathrm{extended}}$ | $T^*$ (K) | $D^*$ (Fixed) (au) | $\alpha$ | $\rho_{b,5}$ (g cm$^{-3}$) | Breaking Radius (au) | $\rho_{(2\,\mathrm{au})}$ (g cm$^{-3}$) |
|---|---|---|---|---|---|---|---|---|---|
| B1-bN | 0.29 | 0.05 | 3.11 | 450 | 4 | −2.9 | $10^{-11.7}$ | 40 | $4.29 \times 10^{-13}$ |
| B1-bS | 0.20 | 2.25 | 4.29 | 550 | 4 | −3.3 | $10^{-10.9}$ | 80 | $3.38 \times 10^{-13}$ |

### 5.3. Mass of B1-bN and B1-bS

The dust and gas mass of the envelopes, $M_{\mathrm{env}}$, in our best models for B1-bN and B1-bS are calculated to be 3.32 and 3.82 $M_\odot$, respectively (Table 10) by integrating the input dust densities from 2 au to 1600 au from our best model using a gas-to-dust ratio of 100. The outer radius of 1600 au is chosen from the radius of 3$\sigma$ detection in the 1.3 mm SMA observation (~5″3 for both B1-bN and B1-BS). Compared with the $M_{\mathrm{env}}$

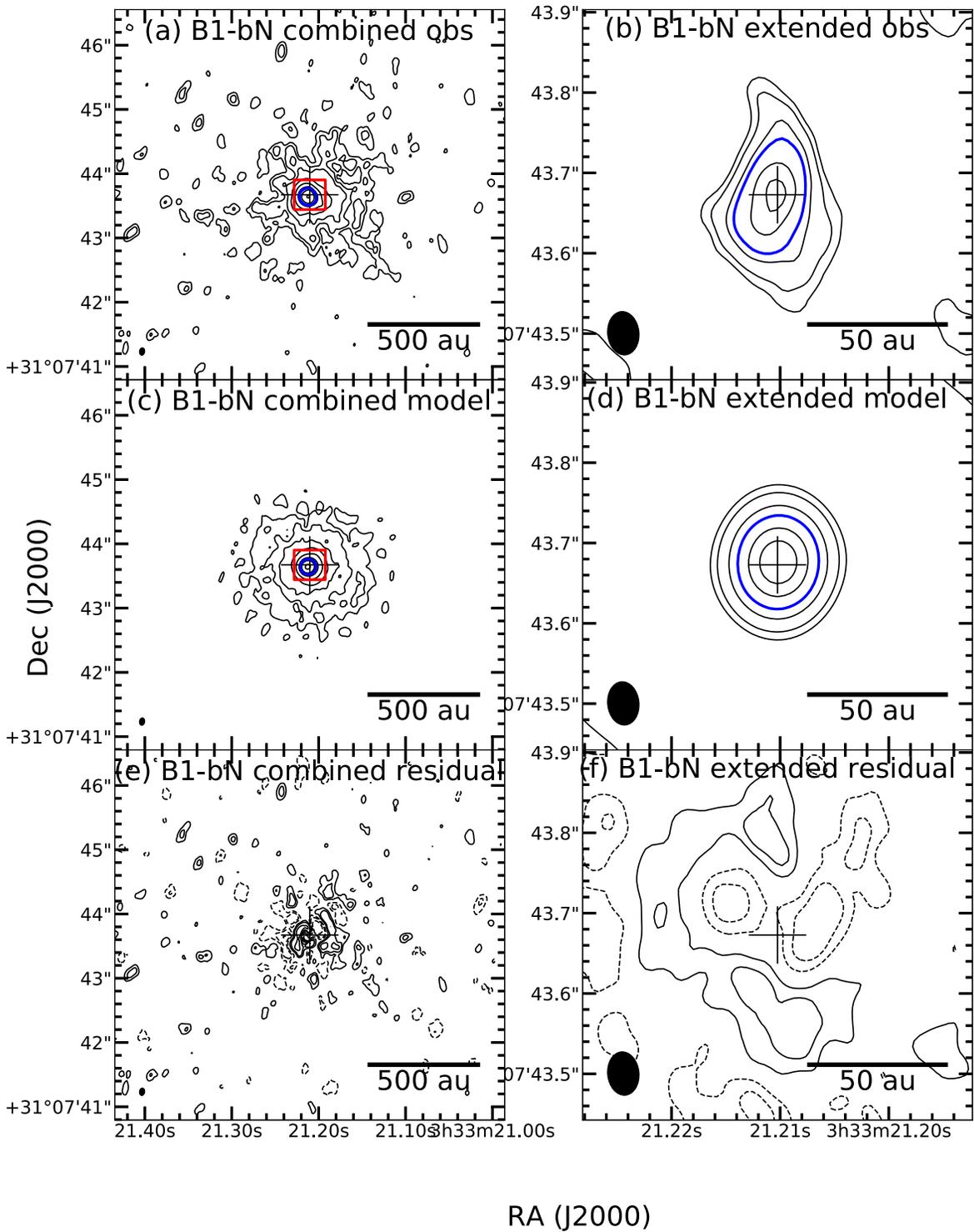

**Figure 5.** ALMA 870 μm intensity maps of observational data ((a) and (b)), models ((c) and (d)), and the residuals of model-subtracted observations ((e) and (f)) for B1-bN with combined and extended configurations. The black contours of (a), (c), and (e) start at $3\sigma$ and are drawn at $6\sigma$, $12\sigma$, $27\sigma$, $54\sigma$, $93\sigma$, and $144\sigma$ of the observational data. The black contours of (b), (d), and (f) start at $3\sigma$ and are drawn at $6\sigma$, $12\sigma$, $24\sigma$, $30\sigma$, and $42\sigma$ of the observational data. The dashed contours of the residual maps are drawn at $-3\sigma$ and $-6\sigma$ of the observational data. The crosses denote the peak positions of B1-bN. The black ellipses show the beam size of each configuration. The blue contours on each panel show the intensity at half of the peak. Red boxes on the combined maps indicate the same field of view of the extended configuration configuration maps.

values derived from the dust emission by Hirano & Liu (2014), 0.36 $M_\odot$ for both sources, our values are higher by a factor of ∼10. There are two reasons for the different mass estimations. First, the adopted distances are different: 230 pc for Hirano & Liu (2014) and 301 pc for our model. The larger distance increases the mass by a factor of 1.7. Second, the dust-mass opacity we used in our radiative transfer calculation, $\kappa_\nu = 0.002$ cm$^2$ g$^{-1}$ at $\nu = 230$ GHz from $R_V = 5.5$, from Weingartner & Draine (2001), is lower than the value used in Hirano & Liu (2014), $\kappa_\nu = 0.01$ cm$^2$ g$^{-1}$ at $\nu = 230$ GHz from

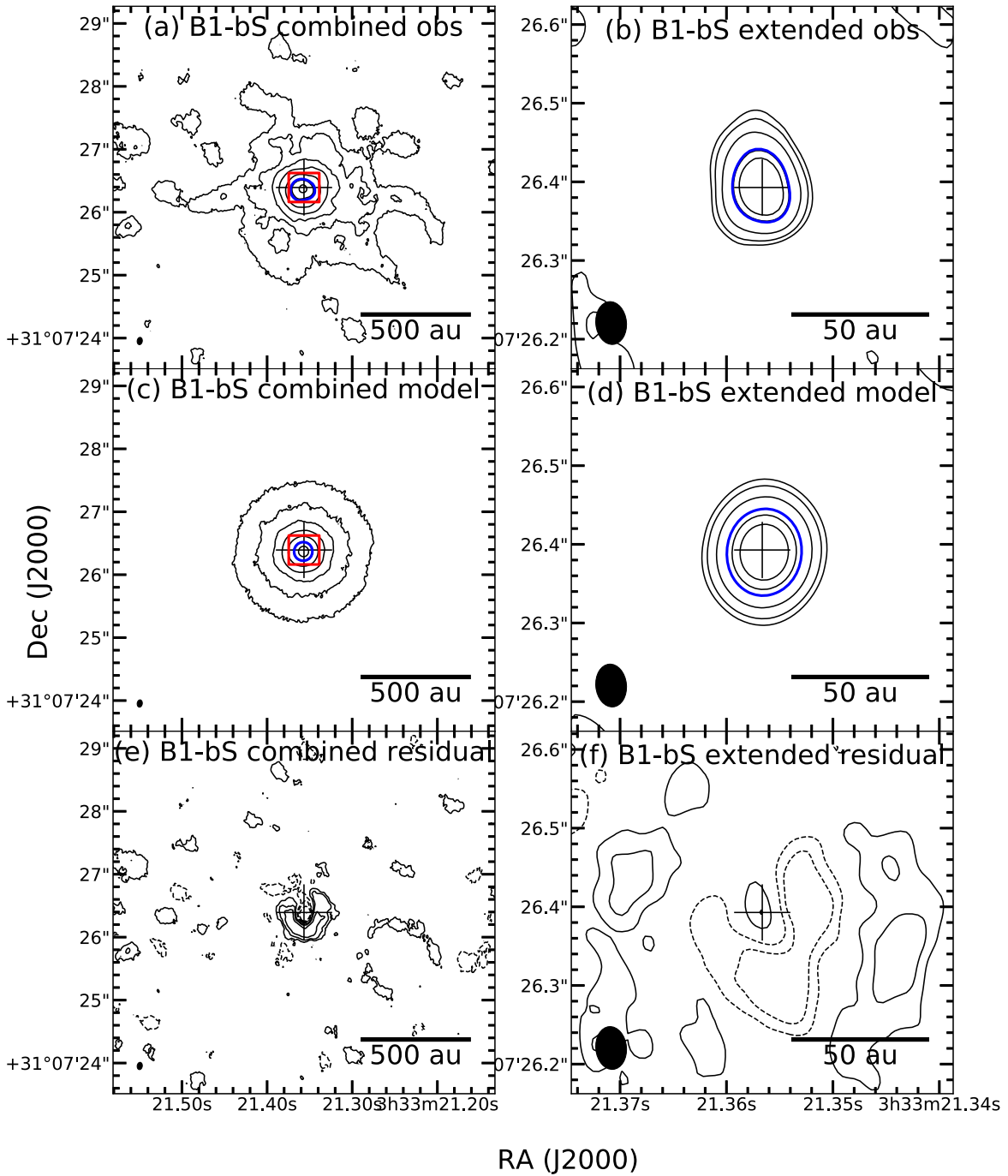

**Figure 6.** ALMA 870 μm intensity maps of observational data ((a) and (b)), models ((c) and (d)), and the residuals of model-subtracted observations ((e) and (f)) for B1-bS with combined and extended configurations. The black contours of (a), (c), and (e) start at 3σ and are drawn at 6σ, 12σ, 27σ, 54σ, 93σ, and 144σ of the observational data. The black contours of (b), (d), and (f) start at 3σ and are drawn at 6σ, 12σ, 24σ, 30σ, and 42σ of the observational data. The dashed contours of the residual maps are drawn at −3σ and −6σ of the observational data. The crosses denote the peak positions of B1-bS. The black ellipses show the beam size of each configuration. The blue contours on each panel show the intensity at half of the peak. Red boxes on the combined maps indicate the same field of view of the extended configuration maps.

Ossenkopf & Henning (1994). The smaller kappa increases the mass by a factor of 5. Thus, the envelope mass of B1-bN and B1-bS are comparable with those derived by Hirano & Liu (2014) if we use the same distance and $\kappa_\nu$. The larger $M_{env}$ of B1-bS compared to B1-bN suggests that B1-bS may have a larger mass reservoir for future accretion onto the first core, thus $M_{env}$ may not be a direct indicator for the evolutionary stages of the two sources. The envelope mass can only affect the final mass of the source.

### 5.4. Density and Temperature Profiles of B1-bN and B1-bS

We compare the density and temperature of our best models with radiation-hydrodynamic simulations of FHCs.

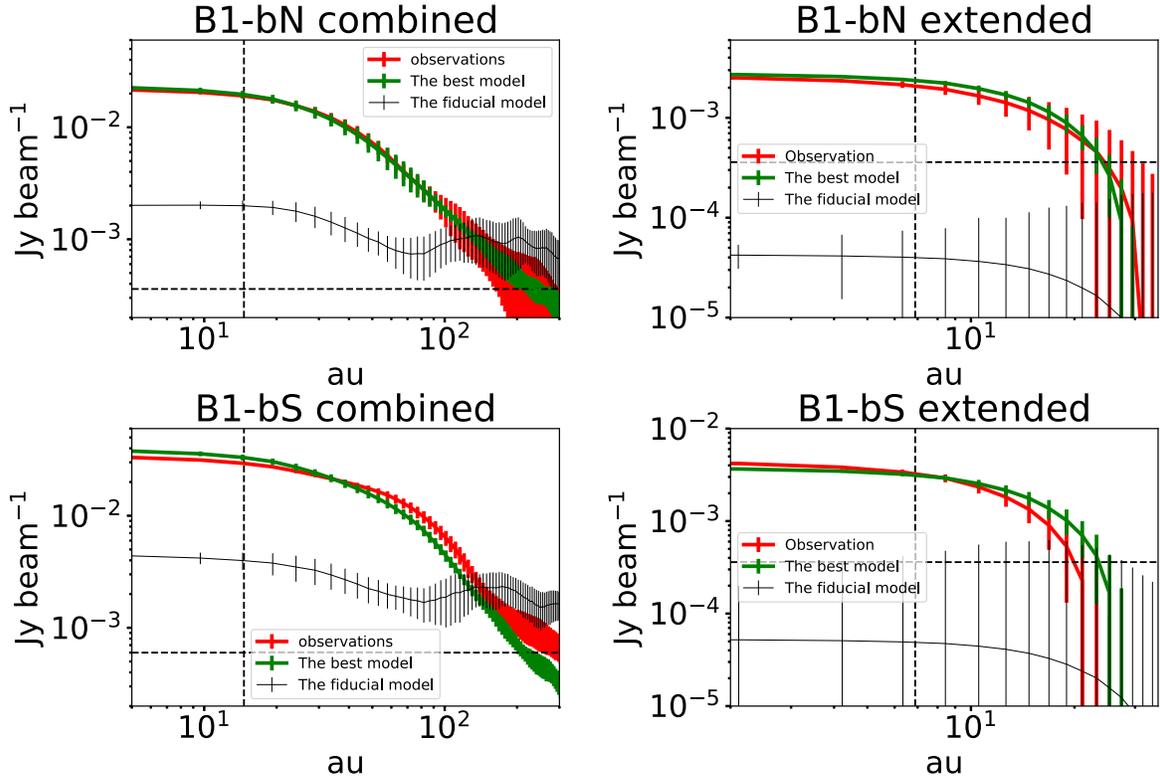

**Figure 7.** Annually averaged intensity profiles of B1-bN and B1-bS. Red lines are the observed profiles. Green lines show the best model profiles. Black lines are the fiducial model. Horizontal dashed lines indicate the $3\sigma$ detection threshold of each observation. Vertical dashed lines indicate the beam size of each configuration.

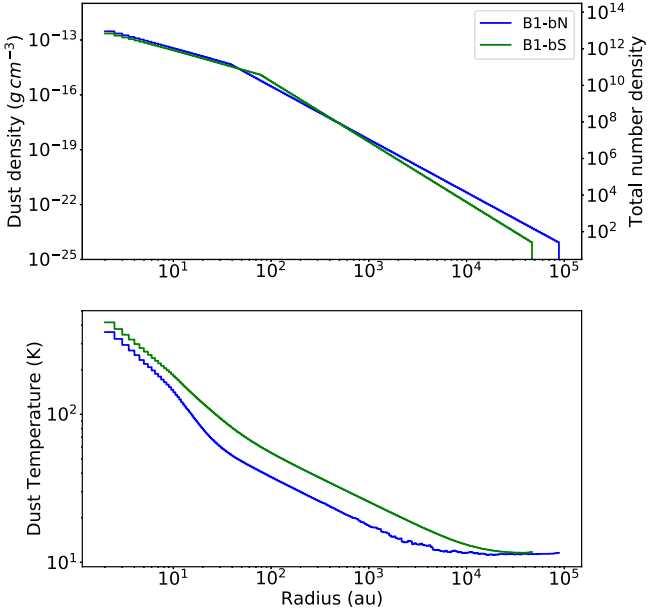

**Figure 8.** The best model envelope dust-density profiles (upper panel) and dust-temperature profiles (lower panel) for B1-bN and B1-bS. The $H_2$ number density is calculated using a gas-to-dust ratio of 100.

Masunaga et al. (1998) provide information of the density and temperature at different stages for a 1 $M_\odot$ cloud-collapse simulation with an initial temperature of 10 K (M1a model). We find that, by comparing the densities ($\rho$(2 au)) and the temperatures of the hot compact sources in our models, B1-bN and B1-bS both fall in the ranges of the FHC stage. We further compare our density and temperature profiles with Bhandare et al. (2018). The paper presents snapshots of density and temperature profiles right at the first collapse and the second collapse for a series of initial cloud masses. We find that B1-bN's and B1-bS's density and temperature profiles both lie between the first collapse and second collapse and close to the second collapse stage (although the slope of the density profiles are different; see Section 5.5 for further discussions). From the comparison above, we show that B1-bN and B1-bS could be in the FHC stage, and close to the second collapse stage. However, in the case of the rotating FHC, the second collapse occurs only at the center of the FHC (e.g., Machida et al. 2008; Saigo et al. 2008). Since the density profile in the remaining part maintains the original profile after the second collapse, it would be difficult to distinguish the FHC before the second collapse and the remnant FHC surrounding the second core.

### 5.5. The Envelope Structure of B1-bN and B1-bS

We compare the breaking radius and the power-law index of our best-fit density profiles with theoretical expectations and observations to investigate the evolutionary stage of the two sources. First, according to the inside-out collapse model, a discontinuity in power-law indices of the density profile separates the inner infalling region from the outer static region, and the discontinuity will propagate outward with sound speed during the collapse (Shu 1977). Our best model density profiles reveal that B1-bS may have gone through the collapsing process earlier compared to B1-bN, due to the larger discontinuity point (breaking radius) found in the density profiles. We calculate the timescale for the expansion wave of the inside-out collapse to propagate to 100 au with a temperature of 10 K is ~1500 yr. This number is shorter than the timescale of an FHC (∼5000 yr; Masunaga & Inutsuka 2000).

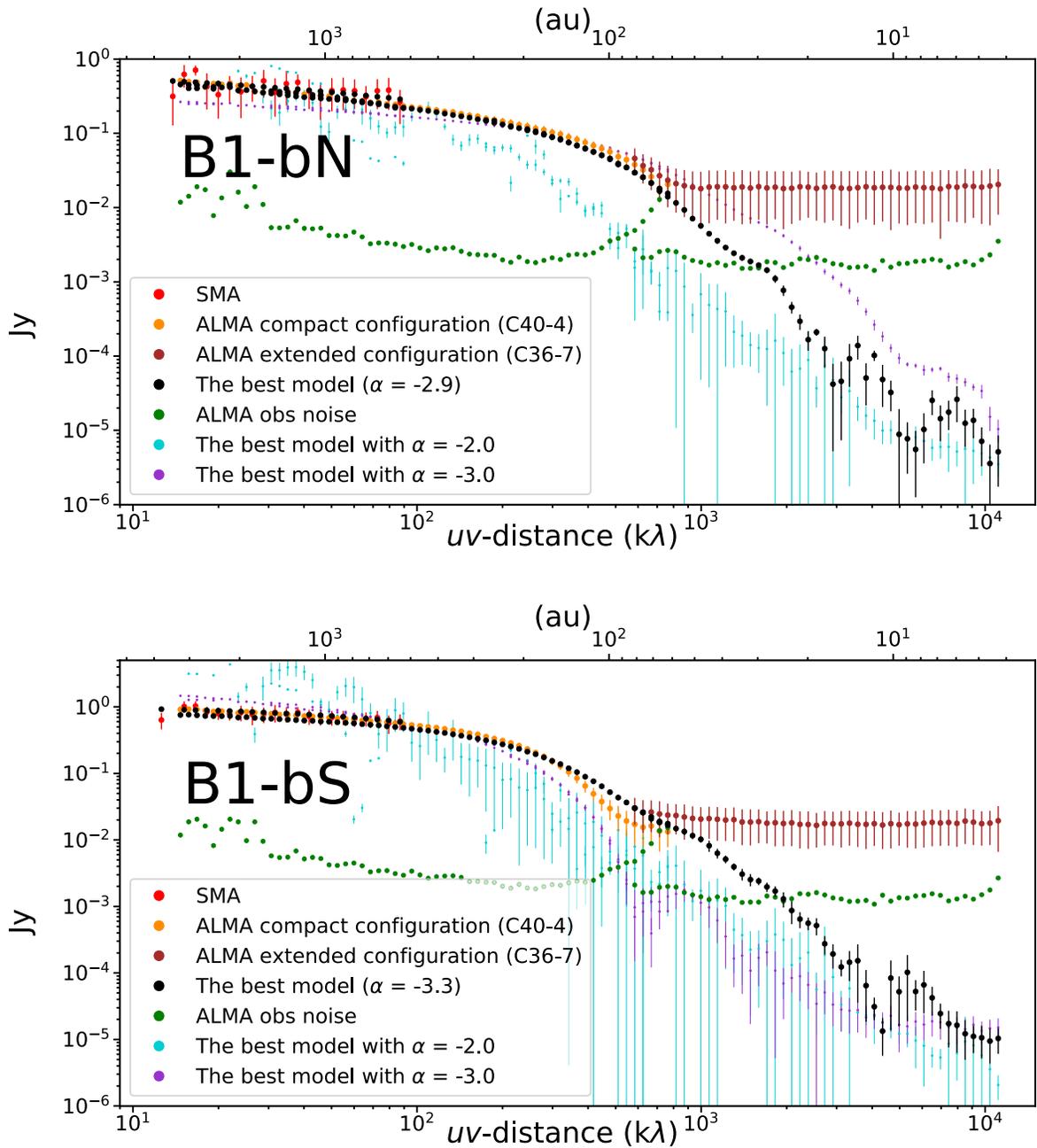

**Figure 9.** Visibilities of observational data and the best model for B1-bN (upper panel) and B1-bS (lower panel) at 870 μm. The data are binned to improve the signal-to-noise ratio. The bin width is 0.028, logarithmically. The observational noise data points are the average of all 1-baseline noise within the bin, the same way the visibilities are averaged. The cyan points show the best model with $\alpha = -2.0$ and the purple points show the best model with $\alpha = -3.0$.

Second, our best model profiles do not follow the envelope density distribution as $\rho \propto r^{-2}$. Masunaga et al. (1998) developed the density distribution from homogeneous to the centrally peaked $r^{-2}$ profile throughout the first collapse phase, as well as the Larson–Penston solution (Larson 1969; Penston 1969). Bhandare et al. (2018) also show $\rho \propto r^{-2}$ at the $r > 10$ au right at at the second collapse. Our results show the power index $\alpha = -2.9$ for B1-bN and $-3.3$ for B1-bS. In Figure 9, we show the visibilities of the best models but with $\alpha = -2.0$ and

$-3.0$. It is clear that the different power indices of the envelope are distinguishable in the visibilities. Vorobyov & Basu (2005) find that, starting in the prestellar runaway collapse phase, a shortage of matter developing at the outer edge of a core generates an inward-propagating rarefaction wave that steepens the radial gas-density profile in the envelope from $\rho \propto r^{-2}$ (self-similar solution) to $\rho \propto r^{-3}$ or even steeper. Observationally, Hung et al. (2010) find that, around starless to Class 0 stages, the power index could be steeper than $-2$. For example, several starless cores show a steep power-law index, such as Serp-Bolo19 ($\alpha = -2.7$) and L694-2 ($\alpha = -2.6$). There are also several examples of Class 0 sources, such as CB 68 ($\alpha = -2.6$) and L483 ($\alpha = -2.5$). These studies support our results showing that the power-law index for the envelope-density profile could be steeper than $\alpha = -2$ for the two sources if they are indeed in the FHC stage.

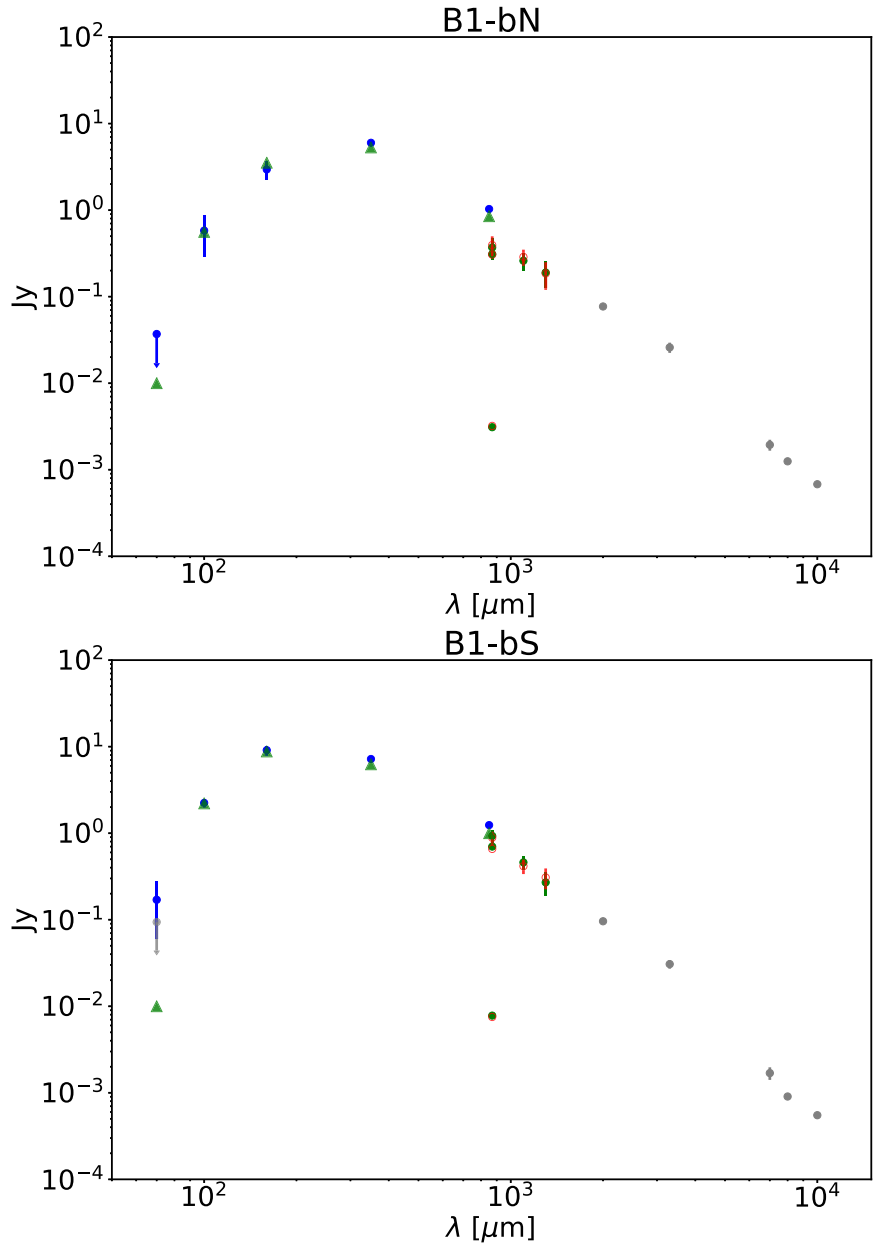

**Figure 10.** Spectral energy distributions of B1-bN (top panel) and B1-bS (bottom panel). Blue points are the observed flux densities from the Herschel PACS 70, 100, and 160 $\mu$m after the color correction (Pezzuto et al. 2012), the CSO SHARC 350 $\mu$m (Hirano & Liu 2014), the JCMT SCUBA 850 $\mu$m (Hirano & Liu 2014), and green triangle points are the flux density from the sky brightness of the best models within the same beam size of these observations. The red circles are the flux densities derived from the interferometric observations that we utilized in this work: the SMA 870 $\mu$m (upper), the combined ALMA 870 $\mu$m (middle), the ALMA 870 $\mu$m with extended configuration (lower), and the SMA 1.1 and 1.3 mm, and green points are the flux densities derived from the best model synthesized images. We also show the Spitzer MIPS 70 $\mu$m (Hirano & Liu 2014), PdBI 2 mm (Gerin et al. 2015), the Nobeyama Millimeter Array 3.3 mm (Hirano & Liu 2014), the Very Large Array (VLA) 7 mm (Hirano & Liu 2014), and the VLA 0.8 and 1.0 cm (Tobin et al. 2016) in gray points. Arrows mark the upper limits.

### 5.6. Evolutionary Stage of B1-bN and B1-bS

We have shown that the physical properties of the central compact objects embedded in B1-bN and B1-bS are consistent with those of FHCs. The density profiles reveal that B1-bS may have gone through the collapsing process earlier than B1-bN. Our conclusion is consistent with recent studies arguing that B1-bS is slightly more evolved due to the detection of $H^{13}CO^+$ (Huang & Hirano 2013) and complex organic molecules (Marcelino et al. 2018; van Gelder & Tabone 2020), higher velocity outflows (Hirano & Liu 2014; Gerin et al. 2015), and a steeper increase in mass with radius than B1-bN (Gerin et al. 2017).

### 6. Conclusion

In this work, we have explored the physical properties of two well-known FHC candidates, B1-bN and B1-bS, by matching

x

**Table 10**
Properties of B1-bN and B1-bS

| Object Name | $L_*$ ($L_\odot$) | $L_{bol}$ ($L_\odot$) | $M_{env}$ ($M_\odot$) |
|---|---|---|---|
| B1-bN | 6.70 | 0.18 | 3.32 |
| B1-bS | 14.97 | 0.26 | 3.82 |

**Note.** $L_*$ is the intrinsic luminosity of the FHC of our best model. $L_{bol}$ is the bolometric luminosity of the source of our best model.

interferometric data from SMA 1.1 and 1.3 mm and ALMA 870 μm observations with simulated synthesis images of the two sources. We produced the images based on a simple model that contains a single hot compact first-core-like component at the center surrounded by a large-scale, cold and dusty envelope, which is described by a broken power-law density distribution. The main conclusions are as follows.

1. Our best models show that, with a fixed size of 4 au, the temperature of the hot compact component of the two first-core candidates are 450 K and for B1-bN and 550 K for B1-bS. These values are consistent with first-core properties suggested by theoretical predictions, which have sizes of ∼5 au and a temperature lower than 2000 K (Masunaga et al. 1998). These results suggest that B1-bN and B1-bS are at an early evolutionary stage.

2. We show that B1-bN and B1-bS could be in the FHC stage by comparing densities and temperatures of our best model with radiation-hydrodynamic simulations of FHCs from Masunaga et al. (1998) and Bhandare et al. (2018).

3. The power-law indices for the envelope-density profiles are $\alpha \sim -2.9$ for B1-bN and $\alpha \sim -3.3$ for B1-bS. These power-law indices are steeper than $\alpha = -2$ in the self-similar solution, which are predicted by theoretical calculations of a collapsing, bound envelope (Vorobyov & Basu 2005) and consistent with observations (Hung et al. 2010).

4. According to the inside-out collapse model (Shu 1977), our best-fit density profiles suggest that B1-bS may have started the collapsing process earlier compared to B1-bN, since a larger discontinuity point (breaking radius) is found in the density profile ($R_b$ at ∼40 au for B1-bN and ∼80 au for B1-bS).

5. Our modeling method shows that the resolutions of ∼10, ∼100, and ∼1000 au scale interferometric observations are all important and indispensable for exploring the physical properties of an FHC candidate. The 1000 au resolution observation can provide assessment of the envelope-density distribution, while the ∼10 to ∼100 au resolution observations are crucial to constraining the breaking radius of the broken power-law density distribution, and accurately deriving the luminosity of the central hot compact first-core-like object.

We have shown that the models of B1-bN and B1-bS are consistent with the FHC in numerical simulation. It is desirable to investigate the evolutionary stages of all other FHC candidates using the same method of this work. The investigation of FHC properties will help make great strides in our understanding of low-mass star formation. Moreover, applying this method to Class 0/I protostars may provide properties of the central heating source and envelope power-law profiles for different stages across a wider timescale.

This work used high-performance computing facilities operated by the Center for Informatics and Computation in Astronomy (CICA) at National Tsing Hua University. This equipment was funded by the Ministry of Education of Taiwan, the Ministry of Science and Technology of Taiwan, and National Tsing Hua University. This paper makes use of the following ALMA data: ADS/JAO.ALMA#2015.1.00025.S and ADS/JAO.ALMA#2015.A.00012.S. ALMA is a partnership of ESO (representing its member states), NSF (USA), and NINS (Japan), together with NRC (Canada), MOST and ASIAA (Taiwan), and KASI (Republic of Korea), in cooperation with the Republic of Chile. The Joint ALMA Observatory is operated by ESO, AUI/NRAO, and NAOJ. H.Y.D. and S.P.L. acknowledge grants from the National Science and Technology Council, R.O.C 106-2119-M-007-021-MY3 and 109-2112-M-007-010-MY3. N.H. acknowledges support from the National Science and Technology Council (NSTC) with grants 110-2112-M-001-048 and 111-2112-M-001-060.

*Software:* CASA (McMullin et al. 2007), MIRIAD (Sault et al. 1995), RADMC-3D (Dullemond et al. 2012b).

### ORCID iDs

Hao-Yuan Duan (段皓元)  https://orcid.org/0000-0002-7022-4742
Shih-Ping Lai (賴詩萍)  https://orcid.org/0000-0001-5522-486X
Naomi Hirano (平野尚美)  https://orcid.org/0000-0001-9304-7884
Travis J. Thieme (陸哲軒)  https://orcid.org/0000-0003-0334-1583